\documentclass[sigconf,natbib=false]{acmart}
\AtBeginDocument{%
  }

\copyrightyear{2025}
\acmYear{2025}
\setcopyright{cc}
\setcctype{by-nc-sa}
\acmConference[PROMISE '25]{21st International Conference on Predictive Models and Data Analytics in Software Engineering}{June 26, 2025}{Trondheim, Norway}
\acmBooktitle{21st International Conference on Predictive Models and Data Analytics in Software Engineering (PROMISE '25), June 26, 2025, Trondheim, Norway}\acmDOI{10.1145/3727582.3728683}
\acmISBN{979-8-4007-1594-5/2025/06}
\usepackage[utf8]{inputenc}

\RequirePackage[
  datamodel=acmdatamodel,
  style=acmnumeric,
  ]{biblatex}

\usepackage{tikz}
\usetikzlibrary{shapes.misc}
\newcommand{\mytimes}{ \tikz[baseline=-.55ex] \node [inner sep=0pt,cross out,draw,line width=1pt,minimum size=1ex] (a) {};}
\usepackage{algorithmic}
\usepackage{graphicx}
\usepackage{textcomp}
\usepackage{xcolor}
\usepackage{dirtree,array}
\usepackage{listings}
\usepackage{booktabs}
\usepackage{multirow}
\usepackage{enumitem}
\usepackage{soul} 
\usepackage{url}

\definecolor{RED}{RGB}{255,0,0}
\definecolor{ORANGE}{RGB}{255,128,0}
\definecolor{GREEN}{RGB}{0,128,0}
\definecolor{GREY}{RGB}{128,128,128}
\definecolor{BLUE}{RGB}{65,105,225}
\definecolor{mediumblue}{RGB}{1,1,1}
\definecolor{PURPLE}{RGB}{160, 32, 240}

\begin{document}

\lstdefinestyle{interfaces}{
  float=tp,
  floatplacement=tbp,
  abovecaptionskip=-5pt
}
\title{A Qualitative Investigation into LLM-Generated Multilingual Code Comments and Automatic Evaluation Metrics}

\author{Jonathan Katzy}
\affiliation{
 \institution{Delft University of Technology}
 \city{Delft}
 \country{The Netherlands}
}
\orcid{0009-0005-9574-2414}
\email{J.B.Katzy@TUDelft.nl}

\author{Yongcheng Huang}
\affiliation{
 \institution{Delft University of Technology}
\city{Delft}
\country{The Netherlands}
}
\orcid{0000-0002-7987-6317}
\email{Y.Huang-51@Student.TUDelft.nl}

\author{Gopal-Raj Panchu}
\affiliation{
 \institution{Delft University of Technology}
\city{Delft}
 \country{The Netherlands}
}
\orcid{0009-0007-1113-7933}
\email{G.G.S.Panchu@Student.TUDelft.nl}

\author{Maksym Ziemlewski}
\affiliation{
 \institution{Delft University of Technology}
 \city{Delft}
 \country{The Netherlands}
}
\orcid{0009-0007-2386-0360}
\email{M.Ziemlewski@Student.TUDelft.nl}

\author{Paris Loizides}
\affiliation{
 \institution{Delft University of Technology}
\city{Delft}
 \country{The Netherlands}
}
\orcid{0009-0009-0461-5305}
\email{P.Loizides@Student.TUDelft.nl}

\author{Sander Vermeulen}
\affiliation{
 \institution{Delft University of Technology}
\city{Delft}
 \country{The Netherlands}
}
\orcid{0009-0008-9355-3611}
\email{S.R.Vermeulen@Student.TUDelft.nl}

\author{Arie van Deursen}
\affiliation{
 \institution{Delft University of Technology}
 \city{Delft}
 \country{The Netherlands}
}
\orcid{0000-0003-4850-3312}
\email{Arie.vanDeursen@TUDelft.nl}

\author{Maliheh Izadi}
\affiliation{
 \institution{Delft University of Technology}
 \city{Delft}
 \country{The Netherlands}
}
\orcid{0000-0001-5093-5523}
\email{M.Izadi@TUDelft.nl}

\renewcommand{\shortauthors}{Katzy et al.}

\begin{abstract}
Large Language Models are essential coding assistants, yet their training is predominantly English-centric. In this study, we evaluate the performance of code language models in non-English contexts, identifying challenges in their adoption and integration into multilingual workflows.
We conduct an open-coding study to analyze errors in code comments generated by five state-of-the-art code models, CodeGemma, CodeLlama, CodeQwen1.5, GraniteCode, and StarCoder2 across five natural languages: Chinese, Dutch, English, Greek, and Polish. Our study yields a dataset of 12,500 labeled generations, which we publicly release. We then assess the reliability of standard metrics in capturing comment \textit{correctness} across languages and evaluate their trustworthiness as judgment criteria.
Through our open-coding investigation, we identified a taxonomy of 26 distinct error categories in model-generated code comments. They highlight variations in language cohesion, informativeness, and syntax adherence across different natural languages. Our analysis shows that, while these models frequently produce partially correct comments, modern neural metrics fail to reliably differentiate meaningful completions from random noise. 
Notably, the significant score overlap between expert-rated correct and incorrect comments calls into question the effectiveness of these metrics in assessing generated comments. 

\end{abstract}





\maketitle

\section{Introduction}
The adoption of Large Language Models (LLMs) for code has led to numerous benefits. When adopting LLMs in the software development process, studies reveal that LLMs boost developer productivity~\cite{ziegler2022productivityassessmentneuralcode}, and facilitate diverse software engineering tasks~\cite{jiang2024survey,izadi2022codefill}. Even in education, LLMs have shown to be beneficial when it comes to teaching programming~\cite{boguslawski2025programming}, and assisting developers of various skill levels~\cite{nam2024using}.

While code models can enhance both learning and development, most research and development in this area focus on \textit{English} contexts. This emphasis limits the adoption of code models for other languages, excluding non-English speakers from the benefits of LLMs in education~\cite{10.1145/3657604.3662036}, access to learning in their preferred language~\cite{language_sweden, 10.1145/2729094.2742618}, and integration of code models into workflows that often involve substantial non-English code~\cite{hu2022practitioners, 7332491}. Beyond adoption challenges, non-English code is sometimes considered a ``data smell'' to be avoided when training code models ~\cite{vitale2024catalog} leading to less and less representation.

By non-English code, we refer to the presence of a language other than English in either the identifiers, literals, or comments. This is often a mixture of English and another language. While non-English code is not commonly used in open source projects, close source proprietary code bases often contain other languages~\cite{7332491}.

\textbf{We aim to evaluate the performance of several state-of-the-art code LLMs in non-English contexts, assess their effectiveness and identify limitations that may impede their adoption in low-resource languages}.
More specifically, we evaluate five code LLMs—CodeGemma~\cite{codegemmateam2024}, CodeLlama~\cite{roziere2023code}, CodeQwen1.5~\cite{codeqwen1.5}, GraniteCode~\cite{mishra2024granitecodemodelsfamily}, and StarCoder2~\cite{lozhkov2024starcoder2stackv2}—across five natural languages: \textit{English}, \textit{Dutch}, \textit{Chinese}, \textit{Greek}, and \textit{Polish}.  
To assess their performance in each language, we focus on comment generation, a task that serves as a proxy for code comprehension and linguistic adaptation.
We take both \textit{qualitative} and \textit{quantitative} approaches to assess the strengths and weaknesses of the selected models.

For the qualitative investigation, we conduct an open-coding investigation into errors and accuracy of our target models. This gives us an overview of errors and insight into the `correctness'  of the comments. Gaining deeper insight into errors made in the generation is essential to understand what the current challenges are with regards to trust~\cite{Witchel2022}, education~\cite{nishanthi2020understanding} and the willingness of practitioners to adopt models into their non-English workflows~\cite{hu2022practitioners}.

For the quantitative evaluation of comment generation, we investigate how well common metrics capture the performance of LLMs in non-English comment generation. We use the expert-labeled comments as a baseline to compare the metrics scores to and analyze their ability to differentiate between correct and incorrect predictions, as well as their ability to differentiate real generations from noise.
More specifically, we answer the following questions; 

\begin{itemize}
    \item What are common errors when LLMs generate comments in different languages?
    \item How well do LLMs generate comments in non-English code?
    \item Do common evaluation metrics give a trustworthy evaluation of predictions?
    \item Do common evaluation metrics effectively differentiate between correct and incorrect predictions?
\end{itemize}

Our investigation resulted in a taxonomy of $26$ distinct errors made by LLMs in $5$ different languages. We show that linguistic errors increase dramatically when working in non-English contexts (up to $15.1\times$), as well as the likelihood that models produce incoherent generations. These increases are higher than the increases in semantic errors, which increased by $3.7\times$ at most. This has strong implications for the adoption of code models into the workflow in non-English settings.

When investigating the metrics, we see that neural metrics struggle to differentiate real predictions from random noise. We showed that there is a substantial overlap between the scores assigned to random noise and to real generations when working with neural metrics. We also see that neural metrics give an overall higher score to all predictions, especially in Chinese, even when human annotators rated most predictions as incorrect.

\section{Related Work}

\paragraph{Multilingual LLMs in NLP}
Studies on LLMs in NLP settings have shown that they perform poorly in non-English and resource-poor natural languages. The performance of several models investigated in Indonesian and African languages continues to fall short, producing responses of lower quality compared to other high-resource languages such as English~\cite{koto2023largelanguagemodelspass, ojo2024goodlargelanguagemodels}. Further studies showed lower performance for non-English queries to chat models resulting in less accurate and lower quality responses compared to English queries, as well as issues arising with regard to cultural nuances when dealing with non-English queries~\cite{zhang2023donttrustchatgptquestion}.
Such findings are consistent with research suggesting a strong bias in multilingual settings when generating code based on non-English instructions. In the case of Code LLaMA, a $37.8\%$ drop in performance (Pass@1 metric) was observed when code generation was prompted in Chinese rather than English~\cite{wang2024exploringmultilingualbiaslarge}. This performance gap emphasizes the underlying bias in LLMs trained predominantly on English-language datasets. 

\paragraph{Comment/Documentation Generation}
Automatic comment generation is a common task that has shown to help developers understand code~\cite{khan2022automaticcodedocumentationgeneration}. 
In this task, language models are used to generate a natural language description of a piece of code. While it has shown to help developer productivity, most studies focus primarily on the performance of models in English code. 
Aside from evaluating models on comment generation, research has shown that for automated comment generation in general; practitioners expect automatically generated comments to be fluent and grammatically correct (82\%), be concise (88\%), and accurate (88\%)~\cite{hu2022practitioners} in order to include them in their workflows.

\paragraph{Qualitative Analyses of LLM Outputs}
Previous works on translation, and specializations on automatic translations have led to the development of Taxonomies like SCATE~\cite{8544345} and evaluation frameworks such as MQM~\cite{mariana2014multidimensional}, used for natural language tasks. These provide a basis for error classification, however they do not capture all intricacies of code related tasks. Further qualitative studies of code generation errors have identified common bugs in LLM-generated code~\cite{tambon2024bugslargelanguagemodels, code4me} as well as given important insights into user preference for model invocations and acceptance rates~\cite{code4me}, however, they make the assumption that code is always written in English, and they focus primarily on the correctness of a program, rather than specific errors in the language.

\paragraph{Non-English Code}

Despite the global prevalence of code written in different languages, non-English code remains significantly under-researched. To the best of our knowledge, only one study has looked at the distribution of non-English language in code repositories~\cite{7332491}. They show that there is a significant increase in the use of German in both comments and identifier names when looking at industry projects, compared to open source projects, which showed to have almost no non-English text present. Datasets that are intended to be used for the training of LLMs, do mention the presence of other languages in the code, however, do not give any more specific information~\cite{lozhkov2024starcoder2stackv2}.

for code LLMs, most research focuses on English contexts; even researchers who are not native English speakers frame their studies within an English paradigm~\cite{LpezNavarro2015WhyDI}, leading to an under-representation of non-English scenarios. Reports from GitHub show that a significant number of non-English speakers contribute code to the platform each year, highlighting the importance of studying code generation in non-English contexts~\cite{octoverse2022}.

\paragraph{Metrics Reliability}
The ability to evaluate the quality of text generated by LLMs is important to fairly evaluate and compare different models. There are, however, some challenge that arise when attempting to evaluate natural language generation programmatically. 
First, many early methods for evaluating textual generations come from the translation domain, where a generation can be compared to multiple reference texts in order to score them appropriately~\cite{papineni2002bleu, lin2004rouge, banerjee2005meteor}. These methods, however, mainly look at an overlap of letters in the target generation with the reference generations. This creates issues when comparing the correctness of the meaning behind the text, which has been proposed to be solved with embedding based metrics~\cite{zhang2019bertscore, zhou2023codebertscore} or model based metrics~\cite{yuan2021bartscore}.
This has led to the development of frameworks that can be used to compare metrics to each other~\cite{xiao2023evaluating} as well as studies measuring the correlation between different metrics~\cite{Fabbri2020SummEvalRS} and their resilience to simple perturbations~\cite{sai-etal-2021-perturbation}. These studies give an overview of what is to be expected from a metric to be a good judge of generated text. We summarize the requirements as follows: A metric must differentiate correct from incorrect samples, agree with human expert expectations, and be able to detect simple perturbations.

\section{Non-English Characteristics}\label{non-english}
To gain an overview of the differences in the selected languages compared to English, we list the defining characteristics and how they may have an influence on the outputs of a model.

\paragraph{Morphosyntactic Complexity}
Morphosyntactic features govern how words change form (morphology) and combine into sentences (syntax). Chinese's uninflected monomorphemic system eliminates grammatical markings like tense/case~\cite{pulleyblank1995outline}, while Polish uses seven grammatical cases and gendered conjugations. Greek combines gendered nouns with flexible word order through inflectional endings, and Dutch employs gendered articles (de/het). These differences add a layer of complexity to the generation of languages compared to English. Furthermore, Polish and Greek are both flexible when it comes to following a Subject Verb Object (SVO) structure of a sentence, making the outputs context dependent~\cite{HaiderSzucsich+2022+1+39, tzanidaki1995greek}

\paragraph{Orthographic Challenges}
Orthographic systems determine how languages represent speech in writing. Greek's technical symbols (e.g., \(\alpha\), \(\beta\)) and diacritics (tonos/diaeresis) coexist with Polish's nine accented characters (ą, ó, etc.), while Chinese uses logograms instead of alphabetic writing. These add a layer of complexity to the tokenization of text, and require language-specific tokens.
Furthermore, these languages face informal script variations: Greeklish (Latinized Greek), diacritic omission in Polish code comments, and Chinese character simplification. These are informal ways of writing a language, often adopted in online spaces which add complexity for the model to learn~\cite{toumazatos-etal-2024-still}. 
Finally, Greek has also been adopted as a common alphabet for use in mathematics, this adds another source of noise that could reduce a model's ability to model Greek.

\paragraph{Semantic Nuances}
Semantic features require understanding the meaning beyond direct translation. Chinese requires contextual interpretation of idioms~\cite{10.1162/tacl_a_00572}, Dutch incorporates English loanwords through prolonged contact~\cite{VERHEIJEN2022100091}, while Polish/Greek use inflection-enabled word order flexibility to convey pragmatic emphasis. Such features demand cultural and syntactic awareness beyond direct translation equivalents, challenging LLMs' ability to preserve intended meaning during code comment generation.

\section{Approach}

\subsection{Data Collection}

\begin{figure}
    \centering
    \includegraphics[width=0.8\linewidth]{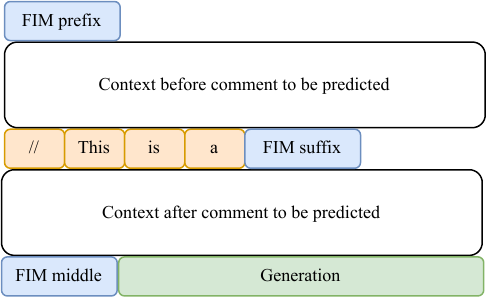}
    \caption{Example input used for inference}
    \label{fig:input_format}
\end{figure}

To gather a code dataset for each natural language, a list of the $2,500$ most common words in each language was used~\footnote{https://github.com/oprogramador/most-common-words-by-language}, based on the OpenSubtitles2016 dataset~\cite{lison-tiedemann-2016-opensubtitles2016}. For each word, the GitHub API was queried to gather 100 files containing the specific word. The collected files were then de-duplicated.

Due to varying architectures, each model has its own maximum token length. To ensure a fair comparison between models, we filter the candidate files by their length after tokenization. We then select only files where all of the context fits within the context of the smallest model (CodeGemma~\cite{codegemmateam2024}, $4,128$ tokens).
To define the inference limit, we calculated the average comment length plus three standard deviations of the ground truth of comments. We filtered out all files where the context length + average length + $3$ standard deviations was greater than $4,128$ tokens to ensure that all files could be correctly predicted by all models.

For each file, the comments were extracted using regular expressions. After extracting the comments, the langdetect toolkit~\cite{ercdidip2022} was used to verify that the comments were written in the appropriate language. The target comments were further filtered manually to remove comments which were references to licenses, contained only the authors' personal/contact information, TODOs, or were auto-generated. From the remaining files, comments with less than $10$ tokens in all tokenizers were filtered out. Finally, $500$ random files were sampled for each language, resulting in $2,500$ generated comments for each language.

\paragraph{Comment Generation}
To control the number of variables to account for, such as finetuning tasks and prompting template, we evaluate base models for their ability to generate comments. To do this we use the Fill In the Middle~\cite{FIM} (FIM) input format. This is the same task that is used to train the base LLMs, giving us the best idea of their core performance. In order to ensure that the models will generate predictions in the correct language, we add the first 3 tokens of the ground truth in order to prime the model for the correct language. Not doing so results in almost exclusively English comment generation. To give an overview of the input format we give an example in Figure~\ref{fig:input_format}.

\subsection{Qualitative Analysis}
\paragraph{Open Coding}
To label the generations, an open coding methodology was applied. An initial set of errors based on previous work~\cite{code4me, 8544345, mariana2014multidimensional} was selected and iteratively improved in multiple rounds of discussion between $6$ authors, for each language there was at least one author that had a native speaker level of understanding of the language to ensure that linguistic nuances, grammar correctness, and semantic accuracy were accurately judged.

An initial subset of $200$ files per language was selected for analysis, similar to the final dataset. The generated comments were manually compared to the originals to classify any possible errors according to the initial hypothetical error taxonomy. The results of this labeling iteration were discussed in meetings between $6$ authors, resulting in adjustments to the error taxonomy.
After each iteration, $6$ authors discussed their findings regarding the error taxonomy and adapted the taxonomy to reflect any changes that needed to be made according to newly identified categories. In the process of adapting the taxonomy, inclusion and exclusion criteria were defined for each category to clarify when an error should be labeled within that category.
This evaluation process was repeated with new data until no further adjustments to the error taxonomy were suggested, resulting in five iterations until the error taxonomy no longer changed. The process of generating a final taxonomy, and labeling all files cost $500$ person-hours.

The resulting error taxonomy was then used for error labeling on our final dataset. This dataset contains $500$ files per language for a total of $2,500$ files, with one inference from each model resulting in $5$ inferences per file. This totals $12,500$ labeled comments.
 
\paragraph{Expert Accuracy}
As well as recording the errors made by a model, we also use an evaluation of a native speaker with programming experience to determine whether a prediction is correct, partially correct, or completely incorrect. In practice, model suggestions for code completion may be accepted, but will be edited afterward to better fit the problem at hand~\cite{10.1145/3613904.3641936, 10.1145/3597503.3608128, code4me}. Including partially correct as a label gives us a more nuanced approach for situations where the models are close to a valid answer but still make minor mistakes. This accounts for developers accepting partially correct completions that they later edit themselves~\cite{code4me}.

We define the categories of correct, partially correct, and incorrect as follows. A correct prediction must include all of the information in the original comment but may also include additional, relevant, information that was omitted. A partially correct comment is a comment where the content is largely correct but there is a small error, such as an incorrect variable name, or spelling mistake. All other comments were considered to be incorrect.

\subsection{Quantitative Analysis}
\begin{figure*}
    \centering
    \includegraphics[width=0.8\linewidth]{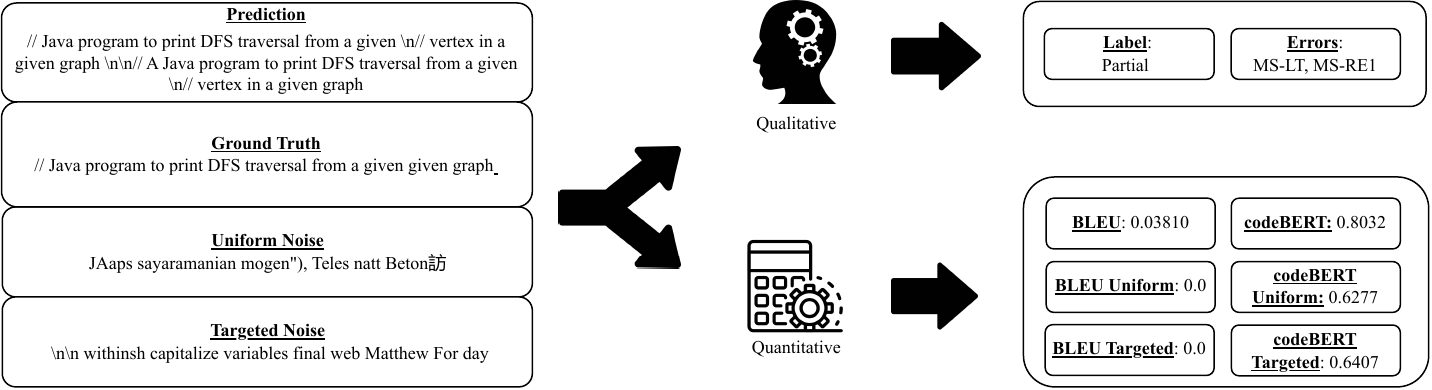}
    \caption{Example Evaluation of a Prediction}
    \label{fig:Example_eval}
\end{figure*}

For the numerical analysis of the comments, and validation of the metrics, we treat the problem as a text generation problem. We have a prediction given by the model and a reference comment we compare it to. For this evaluation we use a combination of word level and neural metrics. To validate the metrics' ability to score a prediction, we focus on the requirement of a metric to reduce the score of a generation significantly due to a perturbation~\cite{sai-etal-2021-perturbation}. We push these perturbations to the extreme. We generate two types of random noise. First we generate ``uniform noise'' which is any token from the models' tokenizer, equal in length to the target comment in tokens, and ``targeted noise'' which are random tokens sampled from the surrounding context. We give an example of the noise and scoring of a generation in Figure~\ref{fig:Example_eval}.

\paragraph{Word Level Metrics}
For the word-level metrics, we use the standard metrics used in code-related studies: BLEU~\cite{papineni2002bleu}, ROUGE~\cite{lin2004rouge}, and METEOR~\cite{banerjee2005meteor}. BLEU measures precision by comparing n-gram matches between the generated and reference summaries while penalizing overgeneration with a brevity penalty. ROUGE focuses on recall by measuring the overlap of n-grams between the reference and the generated texts. ROUGE-1 considers unigrams, ROUGE-2 considers bigrams, and ROUGE-L emphasizes the longest common subsequence to assess fluency and logical consistency. For this investigation, we focus on ROUGE-L.  METEOR combines precision, recall, and synonym-based matching to provide the score. 

\paragraph{Embedding Based Metrics}
These metrics calculate the cosine similarity between embeddings. BERTScore~\cite{zhang2019bertscore} uses a BERT~\cite{kenton2019bert} model to generate the embeddings. The BERT configuration varies according to the target language. For English tasks, a 24-layer RoBERTa large model is used, while for Chinese tasks, a 12-layer BERT Chinese model is utilized. For other languages, it relies on the 12-layer cased multilingual BERT (BERT$_{multi}$) model. BERT$_{multi}$ was trained on 104 languages using the Wikipedia corpora, including Greek, Polish, and Dutch. This model, however, has not been trained on code. CodeBERTScore~\cite{zhou2023codebertscore} is similar to BERTScore, however, it was specifically developed for code evaluation, it differs from BERTScore, by using the CodeBERT model~\cite{feng2020codebert} and ignoring some keywords and syntax tokens common throughout code.

\paragraph{Model Based Metrics} BARTScore~\cite{yuan2021bartscore}
employs BART~\cite{lewis2019bart}, a transformer model that evaluates the quality of generated text through likelihood scores assigned by the model. The BART models is trained by both corrupting the input, and introducing arbitrary noise into the input text, after which it reconstructs the original input.
While specific details regarding training data are not explicitly provided, some indirect information suggests that it includes datasets such as books and Wikipedia entries—similar to those used in RoBERTa's pre-training process~\cite{liu2019roberta}. Similar to BERTScore, code is not specifically mentioned in the training procedure.

\section{Models}
We focus our investigation on open-weight models. This gives us full control over the generations. Ensuring that no other pre-processing steps are performed unknown to us, and allows for reproducible research, which is not guaranteed when using proprietary solutions~\cite{semmelrock2024reproducibilitymachinelearningbasedresearch}. More specifically, we focus on models within the 7B-8B parameter range, the largest amount of parameters that have releases of all $5$ model families. To understand the prevalence of our target languages in the training procedure of these models, we present a brief overview.

\paragraph{CodeQwen1.5-7B}
CodeQwen1.5-7B extends Qwen1.5 with an English and Chinese dataset, supplemented by 3T tokens of code data spanning 92 programming languages, though the distribution of natural languages in this corpus is unspecified~\cite{codeqwen1.5, qwen1.5}.

\paragraph{StarCoder2-7B}
StarCoder2-7B is trained on The Stack V2 dataset\footnote{\url{https://huggingface.co/datasets/bigcode/the-stack-v2}}, totaling 525.5B tokens of programming languages and documentation, with additional natural language data from Wikipedia, StackOverflow, Arxiv, Free Programming Books~\footnote{\url{https://github.com/EbookFoundation/free-programming-books}}, and OpenWebMath, covering all languages in our study~\cite{lozhkov2024starcoder2stackv2}.

\paragraph{Granite-8B-code-base}
Granite-8B-code-base is trained on a collection of programming languages datasets, including GitHub Code Clean\footnote{https://huggingface.co/datasets/codeparrot/github-code-clean}, StarCoderData\footnote{https://huggingface.co/datasets/bigcode/starcoderdata}, and GitHub issues, combined with natural language datasets from StackExchange, Arxiv, and OpenWebMath. English is predominant, with non-English data removed~\cite{mishra2024granitecodemodelsfamily}.

\paragraph{CodeLlama-7B}
CodeLlama-7B builds on Llama2, whose training corpus is 89.7\% English, with minor representation of Chinese, Dutch, and Polish. CodeLlama is further trained on publicly available code, incorporating 8\% natural language data related to code and 7\% from natural language dataset, though specific origins remain undisclosed~\cite{roziere2023code, touvron2023llama}.

\paragraph{CodeGemma-7B}
CodeGemma-7B originates from the Gemma family, trained on 6T tokens inspired by Gemini, which training details remain undisclosed. It is fine-tuned on a dataset consisting of 80\% code and 20\% natural language, predominantly English, with no specified multi-lingual focus. However, dataset sources are also undisclosed~\cite{codegemmateam2024, gemmateam2024gemmaopenmodelsbased}.

\begin{table*}[!ht]
\centering
\begin{minipage}{16cm}
\caption{Taxonomy of common errors found during the open coding process, all languages are listed separately, and summed up in the total. Sub-categories are accumulated in the count of the parent category.}
\begin{tabular}{lr|ccccc}
\toprule
\textbf{Failure category with label ID} & \textbf{Total} & \textbf{Chinese} & \textbf{Dutch} & \textbf{English} & \textbf{Greek} & \textbf{Polish}\\
\midrule
\DTsetlength{0.2em}{0.7em}{0.2em}{0.4pt}{0pt}
\begin{minipage}{6cm}\dirtree{% make the column wider here
.1 \textbf{MS Model-specific Errors}.
.2 MS-IG Incoherent Generation.
.2 MS-CC Copy context.
.2 MS-ME Memorization.
.3 MS-ME1 PII.
.3 MS-ME2 URL.
.3 MS-ME3 Training Data.
.2 MS-ET Early Termination.
.2 MS-LT Late Termination.
.2 MS-RE Repetition.
.3 MS-RE1 Pattern Repetition.
.3 MS-RE2 Verbatim Repetition.
}\end{minipage}

&
\DTsetlength{0pt}{0pt}{0pt}{0pt}{0pt}
\begin{minipage}{0.8cm}\dirtree{% make the column wider here
.1 \rightline{\textbf{5,007}}.
.1 \rightline{259}.
.1 \rightline{991}.
.1 \rightline{422}.
.1 \rightline{236}.
.1 \rightline{114}.
.1 \rightline{72}.
.1 \rightline{164}.
.1 \rightline{2,324}.
.1 \rightline{847}.
.1 \rightline{317}.
.1 \rightline{530}.
}
\end{minipage}
&
\DTsetlength{0pt}{0pt}{0pt}{0pt}{0pt}
\begin{minipage}{1cm}\dirtree{% make the column wider here
.1 \rightline{1,191}.
.1 \rightline{36}.
.1 \rightline{81}.
.1 \rightline{224}.
.1 \rightline{71}.
.1 \rightline{99}.
.1 \rightline{54}.
.1 \rightline{34}.
.1 \rightline{660}.
.1 \rightline{156}.
.1 \rightline{55}.
.1 \rightline{101}.
}
\end{minipage}
&
\DTsetlength{0pt}{0pt}{0pt}{0pt}{0pt}
\begin{minipage}{1cm}\dirtree{% make the column wider here
.1 \rightline{1,072}.
.1 \rightline{42}.
.1 \rightline{291}.
.1 \rightline{36}.
.1 \rightline{32}.
.1 \rightline{1}.
.1 \rightline{3}.
.1 \rightline{47}.
.1 \rightline{405}.
.1 \rightline{251}.
.1 \rightline{109}.
.1 \rightline{142}.
}
\end{minipage}
&
\DTsetlength{0pt}{0pt}{0pt}{0pt}{0pt}
\begin{minipage}{1cm}\dirtree{% make the column wider here
.1 \rightline{752}.
.1 \rightline{2}.
.1 \rightline{70}.
.1 \rightline{74}.
.1 \rightline{68}.
.1 \rightline{5}.
.1 \rightline{1}.
.1 \rightline{13}.
.1 \rightline{491}.
.1 \rightline{102}.
.1 \rightline{79}.
.1 \rightline{23}.
}
\end{minipage}
&
\DTsetlength{0pt}{0pt}{0pt}{0pt}{0pt}
\begin{minipage}{1cm}\dirtree{% make the column wider here
.1 \rightline{952}.
.1 \rightline{167}.
.1 \rightline{307}.
.1 \rightline{41}.
.1 \rightline{34}.
.1 \rightline{3}.
.1 \rightline{4}.
.1 \rightline{51}.
.1 \rightline{296}.
.1 \rightline{90}.
.1 \rightline{6}.
.1 \rightline{84}.
}
\end{minipage}
&
\DTsetlength{0pt}{0pt}{0pt}{0pt}{0pt}
\begin{minipage}{1cm}\dirtree{% make the column wider here
.1 \rightline{1,040}.
.1 \rightline{12}.
.1 \rightline{242}.
.1 \rightline{47}.
.1 \rightline{31}.
.1 \rightline{6}.
.1 \rightline{10}.
.1 \rightline{19}.
.1 \rightline{472}.
.1 \rightline{248}.
.1 \rightline{68}.
.1 \rightline{180}.
}
\end{minipage}
\\
\midrule
\DTsetlength{0.2em}{0.8em}{0.2em}{0.4pt}{0pt}
\begin{minipage}{6cm}
\dirtree{%
.1 \textbf{LG Linguistic Error}.
.2 LG-GR Grammar.
.3 LG-GR1 Plurality. 
.3 LG-GR2 Conjugation.
.3 LG-GR3 Gender.
.3 LG-GR4 Language Syntax.
.3 LG-GR5 Cohesion.
.2 LG-IS Incorrect synonym.
.2 LG-WL Wrong language.
.3 LG-WL1 Undesired translations.
.3 LG-WL2 Incorrect language.
}
\end{minipage}
&
\DTsetlength{0pt}{0pt}{0pt}{0pt}{0pt}
\begin{minipage}{0.8cm}
\dirtree{%
.1 \rightline{\textbf{1,728}}.
.1 \rightline{1,420}.
.1 \rightline{17}.
.1 \rightline{169}.
.1 \rightline{130}.
.1 \rightline{227}.
.1 \rightline{876}.
.1 \rightline{128}.
.1 \rightline{180}.
.1 \rightline{40}.
.1 \rightline{140}.
}
\end{minipage}
&
\DTsetlength{0pt}{0pt}{0pt}{0pt}{0pt}
\begin{minipage}{1cm}
\dirtree{%
.1 \rightline{17}.
.1 \rightline{0}.
.1 \rightline{0}.
.1 \rightline{0}.
.1 \rightline{0}.
.1 \rightline{0}.
.1 \rightline{0}.
.1 \rightline{0}.
.1 \rightline{17}.
.1 \rightline{1}.
.1 \rightline{16}.
}
\end{minipage}
&
\DTsetlength{0pt}{0pt}{0pt}{0pt}{0pt}
\begin{minipage}{1cm}
\dirtree{%
.1 \rightline{224}.
.1 \rightline{118}.
.1 \rightline{1}.
.1 \rightline{8}.
.1 \rightline{23}.
.1 \rightline{25}.
.1 \rightline{60}.
.1 \rightline{0}.
.1 \rightline{106}.
.1 \rightline{9}.
.1 \rightline{97}.
}
\end{minipage}&
\DTsetlength{0pt}{0pt}{0pt}{0pt}{0pt}
\begin{minipage}{1cm}
\dirtree{%
.1 \rightline{66}.
.1 \rightline{65}.
.1 \rightline{0}.
.1 \rightline{0}.
.1 \rightline{0}.
.1 \rightline{42}.
.1 \rightline{23}.
.1 \rightline{0}.
.1 \rightline{1}.
.1 \rightline{0}.
.1 \rightline{1}.
}
\end{minipage}
&
\DTsetlength{0pt}{0pt}{0pt}{0pt}{0pt}
\begin{minipage}{1cm}
\dirtree{%
.1 \rightline{998}.
.1 \rightline{887}.
.1 \rightline{15}.
.1 \rightline{59}.
.1 \rightline{84}.
.1 \rightline{39}.
.1 \rightline{690}.
.1 \rightline{96}.
.1 \rightline{15}.
.1 \rightline{12}.
.1 \rightline{3}.
}
\end{minipage}
&
\DTsetlength{0pt}{0pt}{0pt}{0pt}{0pt}
\begin{minipage}{1cm}
\dirtree{%
.1 \rightline{423}.
.1 \rightline{350}.
.1 \rightline{1}.
.1 \rightline{102}.
.1 \rightline{23}.
.1 \rightline{121}.
.1 \rightline{103}.
.1 \rightline{32}.
.1 \rightline{41}.
.1 \rightline{18}.
.1 \rightline{23}.
}
\end{minipage}
\\
\midrule
\DTsetlength{0.2em}{0.7em}{0.2em}{0.4pt}{0pt}
\begin{minipage}{5.5cm}
\dirtree{%
.1 \textbf{SE Semantic error}.
.2 SE-MD Missing Details.
.2 SE-TS Too Specific.
.2 SE-HA Hallucination.
.3 SE-HA1 Misplaced Facts.
.3 SE-HA2 Out of Context.
.3 SE-HA3 In context.
.2 SE-CS  Code Inclusion.
.3 SE-CS1 Commented code.
.3 SE-CS2 Runnable code.
.2 SE-OI Omitted Identifier.
}
\end{minipage}
&
\DTsetlength{0pt}{0pt}{0pt}{0pt}{0pt}
\begin{minipage}{0.8cm}
\dirtree{%
.1 \rightline{\textbf{8,333}}.
.1 \rightline{837}.
.1 \rightline{168}.
.1 \rightline{4,239}.
.1 \rightline{325}.
.1 \rightline{550}.
.1 \rightline{3,364}.
.1 \rightline{3,009}.
.1 \rightline{171}.
.1 \rightline{2,838}.
.1 \rightline{80}.
}
\end{minipage}
&
\DTsetlength{0pt}{0pt}{0pt}{0pt}{0pt}
\begin{minipage}{1cm}
\dirtree{%
.1 \rightline{2,915}.
.1 \rightline{413}.
.1 \rightline{91}.
.1 \rightline{1,639}.
.1 \rightline{79}.
.1 \rightline{385}.
.1 \rightline{1,175}.
.1 \rightline{718}.
.1 \rightline{6}.
.1 \rightline{712}.
.1 \rightline{54}.
}
\end{minipage}
&
\DTsetlength{0pt}{5pt}{0pt}{0pt}{0pt}
\begin{minipage}{1cm}
\dirtree{%
.1 \rightline{1,350}.
.1 \rightline{107}.
.1 \rightline{21}.
.1 \rightline{636}.
.1 \rightline{105}.
.1 \rightline{80}.
.1 \rightline{451}.
.1 \rightline{574}.
.1 \rightline{68}.
.1 \rightline{506}.
.1 \rightline{12}.
}
\end{minipage}
&
\DTsetlength{0pt}{0pt}{0pt}{0pt}{0pt}
\begin{minipage}{1cm}
\dirtree{%
.1 \rightline{786}.
.1 \rightline{19}.
.1 \rightline{1}.
.1 \rightline{406}.
.1 \rightline{42}.
.1 \rightline{12}.
.1 \rightline{352}.
.1 \rightline{355}.
.1 \rightline{12}.
.1 \rightline{343}.
.1 \rightline{5}.
}
\end{minipage}
&
\DTsetlength{0pt}{0pt}{0pt}{0pt}{0pt}
\begin{minipage}{1cm}
\dirtree{%
.1 \rightline{1,759}.
.1 \rightline{96}.
.1 \rightline{13}.
.1 \rightline{928}.
.1 \rightline{8}.
.1 \rightline{57}.
.1 \rightline{863}.
.1 \rightline{722}.
.1 \rightline{15}.
.1 \rightline{707}.
.1 \rightline{0}.
}
\end{minipage}
&
\DTsetlength{0pt}{0pt}{0pt}{0pt}{0pt}
\begin{minipage}{1cm}
\dirtree{%
.1 \rightline{1,523}.
.1 \rightline{202}.
.1 \rightline{42}.
.1 \rightline{630}.
.1 \rightline{91}.
.1 \rightline{16}.
.1 \rightline{523}.
.1 \rightline{640}.
.1 \rightline{70}.
.1 \rightline{570}.
.1 \rightline{9}.
}
\end{minipage}
\\
\midrule
\DTsetlength{0.2em}{0.7em}{0.2em}{0.4pt}{0pt}
\begin{minipage}{6cm}
\dirtree{%
.1 \textbf{ST Syntax}.
.2 ST-IF Incorrect comment format.
}
\end{minipage}
&
\DTsetlength{0pt}{0pt}{0pt}{0pt}{0pt}
\begin{minipage}{0.8cm}
\dirtree{%
.1 \rightline{\textbf{84}}.
.1 \rightline{84}.
}
\end{minipage}
&
\DTsetlength{0pt}{0pt}{0pt}{0pt}{0pt}
\begin{minipage}{1cm}
\dirtree{%
.1 \rightline{31}.
.1 \rightline{31}.
}
\end{minipage}
&
\DTsetlength{0pt}{0pt}{0pt}{0pt}{0pt}
\begin{minipage}{1cm}
\dirtree{%
.1 \rightline{2}.
.1 \rightline{2}.
}
\end{minipage}&
\DTsetlength{0pt}{0pt}{0pt}{0pt}{0pt}
\begin{minipage}{1cm}
\dirtree{%
.1 \rightline{21}.
.1 \rightline{21}.
}
\end{minipage}
&
\DTsetlength{0pt}{0pt}{0pt}{0pt}{0pt}
\begin{minipage}{1cm}
\dirtree{%
.1 \rightline{5}.
.1 \rightline{5}.
}
\end{minipage}
&
\DTsetlength{0pt}{0pt}{0pt}{0pt}{0pt}
\begin{minipage}{1cm}
\dirtree{%
.1 \rightline{25}.
.1 \rightline{25}.
}
\end{minipage}
\\

\bottomrule
\end{tabular}
\label{tab:taxonomy}
\end{minipage}
\end{table*}

\section{Results}
\subsection{Error Taxonomy}
From the open coding investigation, we arrive at $4$ main categories of error in comment generation. We present these categories and all their subcategories in Table~\ref{tab:taxonomy}. We use this taxonomy to answer RQ1, What are common errors when LLMs generate comments in different languages? Semantic errors, the most common category, refer to errors that relate to the meaning of the comment. This can relate to the amount of detail provided to the model, whether the model is hallucinating, and whether code is included in/after the comment. Linguistic errors refer to errors made in the language itself. This can refer to grammatical errors, but also to which synonyms the model uses and whether it responds in the correct language. Model-specific errors are errors due to the behavior of the LLMs; this focuses on  memorization, copying information from the context, and knowing when to stop.

Each of the four categories can be divided into subcategories. The most common in the Model-Specific category is the Late Termination (MS-LT) error, which occurs when the model continues to produce additional content after the relevant part of the comment has been completed. On the other hand, there is Early Termination (MS-ET), which means that the prediction stops when the comment is not complete. Other error types are Copy Context (MS-CC) when the model copies the surrounding context verbatim, and Repetition (MS-RE), when the model repeats what it has already generated, this can happen in two ways. The model can repeat a pattern, but make minor changes such as increasing a number (MS-RE1) or the model can repeat a token or set of tokens exactly the same (MS-RE2). Incoherent Generation (MS-IG) occurs when the generation consists of random words or symbols with no logic between them. Memorization (MS-ME) is assigned when the model returns personally identifiable information (MS-ME1), URL (MS-ME2), or the prediction is the same as the erroneous ground truth (MS-ME3), indicating that the comment was in the training set (e.g. the original comment contained a grammatical error and a model repeated that, or the original comment did not refer to the code correctly and the model predicted exactly the same). 

Linguistic errors are divided into three subcategories. The first one is Grammar (LG-GR), where all grammatical errors are assigned. The second one, Incorrect Synonym (LG-IS), occurs when a model uses a similar word with an incorrect meaning in context. The last category is the Wrong Language (LG-WL) category. Due to the use of English loan words in many languages, the models will attempt to translate them, when they should not (LG-WL1). Similarly, the model can also generate most of, or a significant part of a comment in a language other than the target language (LG-WL2). 

In the Semantic group, we distinguish several errors, such as the prediction with Missing Details (SE-MD) or being Too Specific (SE-TS). Many predictions also include code, either executable (SE-CS2) or commented out (SE-CS1). Subcategory for Hallucination generations is divided into three types of errors. Misplaced Facts (SE-HA1) occurs, when random facts that do not align with the expected content are present. Out of Context (SE-HA2), which is a hallucination not grounded in the provided context. and In Context (SE-HA3) Hallucinations, which are educated guesses based on the provided context. Finally, we have the Syntax errors (ST), which are all errors related to the syntax of the comments.
\begin{table*}
\centering
\caption[Expert accuracy of all models and languages]{Expert accuracy for all models and languages ($\checkmark$ correct, $\thicksim$ partial, $\mytimes$ incorrect)}
\label{tab:expert_accuracy}
\begin{tabular}{c|rrr|rrr|rrr|rrr|rrr}
\multicolumn{1}{l}{} & \multicolumn{3}{c}{CodeGemma} & \multicolumn{3}{c}{CodeLlama} & \multicolumn{3}{c}{CodeQwen 1.5} & \multicolumn{3}{c}{GraniteCode} & \multicolumn{3}{c}{StarCoder2} \\
\multicolumn{1}{l|}{} & \multicolumn{1}{c}{$\checkmark$} & \multicolumn{1}{c}{$\thicksim$} & \multicolumn{1}{c|}{$\mytimes$} & \multicolumn{1}{c}{$\checkmark$} & \multicolumn{1}{c}{$\thicksim$} & \multicolumn{1}{c|}{$\mytimes$} & \multicolumn{1}{c}{$\checkmark$} & \multicolumn{1}{c}{$\thicksim$} & \multicolumn{1}{c|}{$\mytimes$} & \multicolumn{1}{c}{$\checkmark$} & \multicolumn{1}{c}{$\thicksim$} & \multicolumn{1}{c|}{$\mytimes$} & \multicolumn{1}{c}{$\checkmark$} & \multicolumn{1}{c}{$\thicksim$} & \multicolumn{1}{c}{$\mytimes$} \\
\hline
Chinese & 55 & 221 & 224 & 63 & 201 & 236 & 84 & 200 & 216 & 67 & 175 & 258 & 20 & 218 & 262 \\
Dutch & 298 & 98 & 104 & 246 & 95 & 159 & 158 & 109 & 233 & 253 & 80 & 167 & 215 & 107 & 178 \\
English & 406 & 76 & 18 & 266 & 162 & 72 & 268 & 183 & 49 & 339 & 123 & 38 & 320 & 145 & 35 \\
Greek & 259 & 117 & 124 & 118 & 119 & 263 & 78 & 165 & 257 & 132 & 130 & 238 & 118 & 131 & 251 \\
Polish & 271 & 146 & 83 & 176 & 164 & 160 & 159 & 157 & 184 & 205 & 151 & 144 & 167 & 154 & 179 \\
\end{tabular}
\end{table*}
\begin{figure*}
    \centering
    \includegraphics[width=0.8\linewidth]{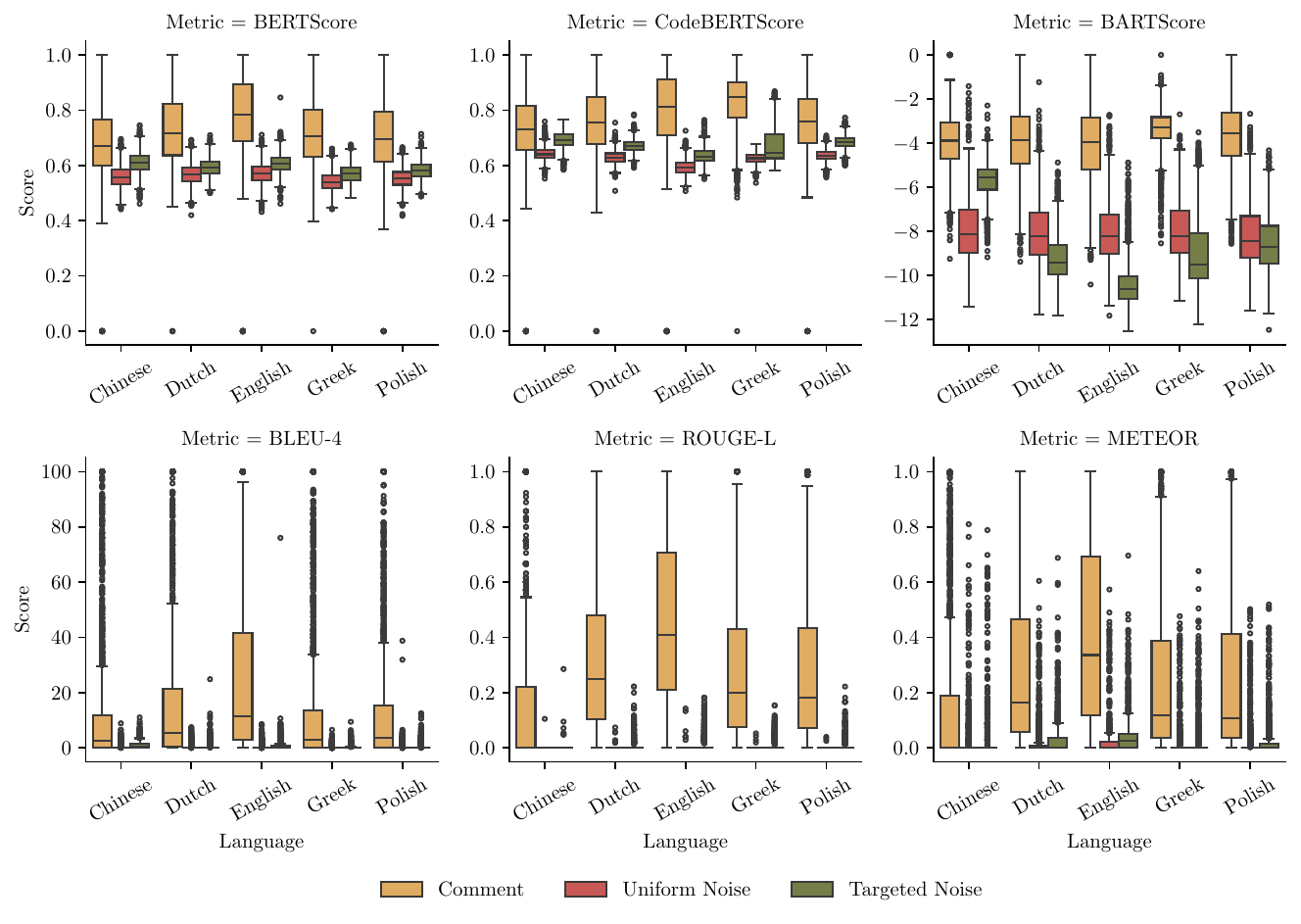}
    \caption{Metric scores comparing LLM generated comments, to random samples of tokens using two separate distributions.}
    \label{fig:metrics}
\end{figure*}

This taxonomy shows that the main issue with comments generated by LLMs is Semantic Errors, often in the form of Hallucinations (SE-HA). These errors account for the largest increase in errors for non-English languages compared to English. Furthermore, we see that the level of detail (SE-MD and SE-TS) is a cause of errors mainly in non-English generations. Looking at the Model Specific errors (MS) we see that for errors related to Memorization (MS-ME), the models are most likely to make them in English compared to other Western languages. Interestingly, Chinese has the highest rate of memorization for all categories. For language-specific errors (LG), we see that Chinese has no grammar errors, as discussed in Section~\ref{non-english}. Furthermore, we notice that Dutch has the highest likelihood of returning predictions in the wrong language. We believe this is due to the similarity of Dutch to English, and the high amount of overlap between the two dictionaries. We also observe a high rate of incorrect synonyms used in Polish compared to all other languages and a high rate of language syntax errors for English, compared to all other language errors made by the model. Finally, in the Syntax errors (ST) we see that the models' ability to adhere to comment syntax is dependent on the language it is using.

We further compare languages on a per-language basis and see that Greek has the highest number of errors for the language grammar, as well as the use of incorrect synonyms. This amounts to an increase of $15.1\times$ the error rate compared to English. This also appears in the MS category, where Greek has the highest tendency to generate incoherent outputs.
On the other hand, Greek is the language that is least likely to generate repetitions out of all analyzed languages. We also see that Dutch, Greek, and Polish are more likely to copy the surrounding context into the comment, while Chinese does it at a rate similar to English.

\begin{figure*}
    \centering
    \includegraphics[width=0.8\linewidth]{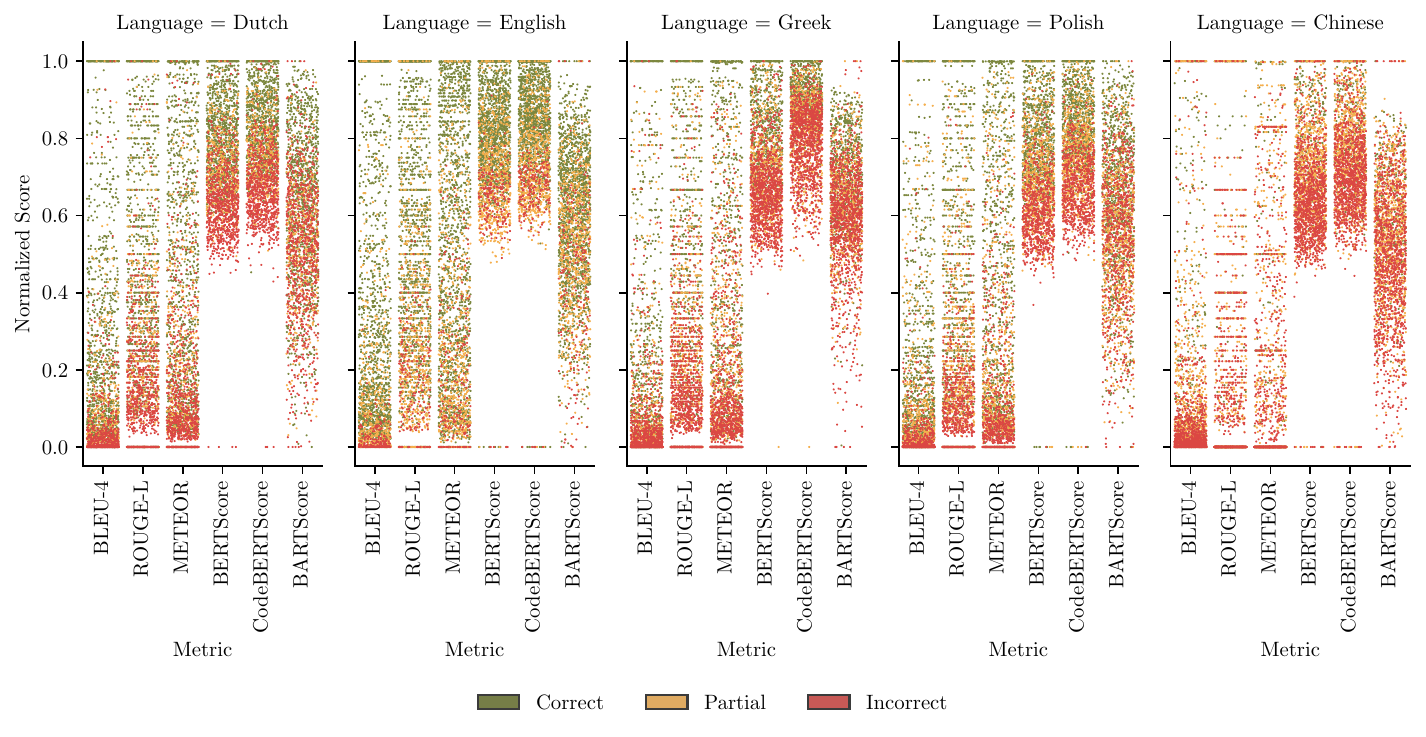}
    \caption{Strip plot, showing the scores assigned to comment generations by different metrics.}
    \label{fig:neural_eval}
\end{figure*}

\subsection{Expert Accuracy}
To understand how well the chosen models perform at generating code comments in languages they were not intended to be used for (RQ2), the experts labeled each generation as being either ``Correct'', ``Partially Correct'', or ``Incorrect''. We give an overview of the scores in Table~\ref{tab:expert_accuracy}. Here we see that while models claim to be only intended to be used for English, and sometimes Chinese, we see that they generate at least partially accurate completions in more than half of the cases for Dutch, Greek, and Polish. Interestingly, we see that although Chinese is the second most common language present in training corpora, it is the worst-performing language overall. On a model level, we see that most models follow the same trend, they perform best in English, then Dutch or Polish, followed by Greek, and finally Chinese. However, for CodeGemma, we see that the performance of Greek is doubled in comparison to other models.

\subsection{Scoring Collections}
We answer RQ3 by analyzing the scores the selected metrics assign to both predictions and the random noise we generated. An overview is given in Figure~\ref{fig:metrics}. Here we look at two factors identified as belonging to a good metric, alignment with expert evaluations, and the ability to differentiate between noise and a real prediction.

Looking at only the English results, we see that the scores assigned by the embedding based metrics align closely with the expert evaluation (around $0.8$). However, when looking at other languages, the embedding-based metrics score the predictions slightly lower, even when the expert evaluation scores the generations much lower (especially for Chinese). The word-based metrics on the other hand score predictions lower in general, however, do match the differences in performance observed in the expert accuracy.

Finally, when looking at noise, we see that for all neural metrics, there is a significant overlap between the scores assigned to noise and the scores assigned to real generations. The BARTScores have a lower overlap compared to the embedding based metrics, however, interestingly, the targeted noise performed worse than the uniform noise for all languages except Chinese. Finally, we see a general trend of CodeBERTScore scoring predictions higher than BERTScore, hinting that the training procedure is a more important factor in a higher BERTScore than generations being studied. For classic metrics, there is a far lower overlap between noise and real comments, and most noise is scored close to $0$ with the METEOR score showing the largest number of outliers towards high scores.

\subsection{Differentiating Correct From Incorrect}
To answer RQ4, Do common evaluation metrics effectively differentiate between correct and incorrect predictions? We plot the scores of selected metrics in Figure~\ref{fig:neural_eval}. Here we show a strip plot of the score for each generated comment and color it by the expert evaluation ``Correct'' (green), ``Partial'' (orange), and ``Incorrect'' (red). Each point in a stripe represents a single generation. To demonstrate the spreads of different scores, we normalize all scores between $0$ and $1$. This shows us if there is a clear separation between the three colors at a glance.
We see a difference in the distribution between neural models, and classical models. Classical models have the majority of their data points scored low and have a tail of scores that score high. On the other hand, neural metrics, especially the embedding-based metrics, have their scores concentrated towards the top, with few points in the lower half of the plot. Looking at all the strip plots, we see no clear separation between correct, partially correct, or incorrect predictions. While there is a slight trend to score correct predictions higher, a large overlap between the three colors persists, showing that the metrics do not separate them accurately.

\section{Discussion}
The results show two main trends. First, we see that the performance of non-English comment generation is often acceptable. However, the generated errors have strong implications for their use by practitioners. Secondly, using neural metrics for automatic evaluation gives an inflated view of an LLM's performance.

We show that the number of errors relating to factuality of the comments increased by at most a factor of $3$ compared to English, the number of errors related to language increased by a factor of $15.1$. This larger decrease in the models' ability to generate accurate natural languages damages the trust users have in the model, and is one of the main requirements practitioners name for automatic comment generation~\cite{hu2022practitioners}. Furthermore, the level of detail (SE-MD) drops significantly (up to $21.8\times$ for Chinese) which is another factor limiting the adoption of models by practitioners. 

Looking at the results of the metric evaluations, we see a worrying trend for neural metrics. The inability of a metric to differentiate random noise from a real generation makes the other evaluations they give questionable. Comparing the expert accuracy scores with the results of the metrics we see that the human evaluators rated English comments correct $5.5\times$ more often than Chinese comments. However, for the neural metrics, the Chinese score is marginally lower than the English score. More worryingly, For CodeBERTScore and BARTScore, the models score Greek better than all other languages, despite being second to last for human raters. Looking at the scores calculated by traditional metrics, we see that while the scores are lower overall, the respective ranking between the different languages matches more closely with the expert evaluators.

Our findings highlight several recommendations for improving multilingual code model development. First, model developers should prioritize diversifying training data to include more non-English code, particularly in languages like Chinese where performance degradation is most severe. Second, evaluation frameworks should require manual checks, as the results demonstrate that neural metrics fail to reliably capture human judgments of comment quality across languages. We recommend maintaining human evaluators in assessment pipelines, particularly when evaluating non-English outputs, until more reliable automated metrics can be developed. Practitioners working with these models should be careful and conduct verification steps when generating comments or documentation in non-English languages. In addition, there is a need to quantify the distribution of non-English code in real-world development scenarios to gain more insight into the impacts of the problem of non-English code generation, and to assess the need to align model training for practical environments.

\section{Threats to Validity}
\paragraph{Memorization Risk}
Most of the models under investigation do not publish their training data. Therefore, it cannot be known whether memorization took place and influenced the answers of the models for the given tasks.
We are unable to control for this due to the prevalence of non-disclosed training datasets. In situations where the memorization was very obvious, we marked it as an error category. We implore open-source models to be released with their training data so research can be conducted fairly.

\paragraph{Labeling Bias}
This study ran the risk of labeling bias due to a limited amount of experts labeling errors per language. We mitigate this bias by defining clear inclusion criteria per error label and having iterative discussions regarding these classifications between six authors. Labeling bias remains a risk in a multilingual context despite our mitigation efforts; therefore, we release all labeled data for validation to the public.

\paragraph{Limited Model Size}
This investigation focused on models in the 7B-8B parameter range. While larger models tend to give better outputs, we limited the model size to the largest size available across all architectures under investigation. This prevents us from including confounding factors; which skew the results of the investigation.

\section{Future Work}
Our study evaluates the performance of five popular models for comment generation in multiple languages and highlights opportunities for future research.

Primarily, we observed a gap in the investigation of the multilingual nature of code. Although it is recognized within the community that code can be written in non-English languages, the extent to which this occurs on a large scale remains largely unknown. Additionally, there is limited research concerning the distribution of non-English language elements within code; these may appear in comments, function names, variable names, or strings, and often coincide with English to varying degrees. This complexity renders the detection of such languages a challenging problem. We propose future research focuses on mapping and annotating other languages present in extensive training datasets for code LLMs.

In our investigation, we also found that neural methods tend to struggle when distinguishing between correct and incorrect predictions. Furthermore, comparisons based solely on average values can be misleading when compared with random noise. These results closely resemble other works on multilingual embeddings of code tokens~\cite{katzy2023impact}. We believe that changing the training objectives to include negative sampling, comparable to word2vec~\cite{mikolov2013efficientestimationwordrepresentations} is a promising avenue for future work. 

Recent research on employing LLMs to replace manual annotation of software engineering artifacts~\cite{ahmed2024llmsreplacemanualannotation} has shown that 
while the annotation generated by LLMs can align with human annotations, they are costly to be evaluated.
To aid in the development and evaluation of these systems, we invite the community to use our publicly available dataset to compare the model outputs with human experts' annotations.

\section{Conclusion}
We explored the ability of LLMs to generate comments in Chinese, Dutch, English, Greek, and Polish. We performed open-coding to create a taxonomy of model failures when generating comments. Expert evaluators assessed the correctness of these comments and analyzed the effectiveness of various metrics in distinguishing noise from real comments, as well as correct from incorrect ones. 

Our qualitative analysis revealed notable behaviors across languages, highlighting which areas are most important to address when designing systems for non-English contexts. Our metric analysis found shortcomings in neural metrics: cosine similarity-based metrics struggled to differentiate noise from genuine predictions, and all neural metrics failed to reliably separate correct from incorrect comments.
To support further research, we release our $12,500$ labeled predictions including five languages and five models~\footnote{https://huggingface.co/datasets/AISE-TUDelft/multilingual-code-comments}.

%%%%%%%%%%%%%%%%%%%%%%%%%%%%%%%
\printbibliography
\end{document}